\documentstyle[prb,aps]{revtex}

\begin{document}
\twocolumn[\hsize\textwidth\columnwidth\hsize\csname
@twocolumnfalse\endcsname
\draft
\preprint{HEP/123-qed}
\title{Magnetic Resonance in the Spin-Peierls compound $\alpha$'-NaV$_{2}$O$_{5}$}
\author{S. Schmidt, W. Palme, B. L\"{u}thi}
\address{Physikalisches Institut, Universit\"{a}t Frankfurt, Robert Mayer Str. 2-4, D-60054 Frankfurt}
\author{M.Weiden, R.Hauptmann and C.Geibel}
\address{Technische Physik, TH Darmstadt, Hochschulstr.8, D-64289 Darmstadt.}
\date{October 24, 1997}
\maketitle
\begin{abstract}
We present results from magnetic resonance measurements for 75-350~GHz in $\alpha$'-NaV$_{2}$O$_{5}$.
The temperature dependence of the integrated intensity indicates that we observe transitions in the excited state.
A quantitative description gives resonances in the triplet state at high symmetry points of the excitation spectrum
of this Spin-Peierls compound. This energy has the same temperature dependence as the Spin-Peierls gap.
Similarities and differences with the other inorganic compound CuGeO$_{3}$ are discussed.
\\
\end{abstract}

]
\narrowtext

Recently the compound $\alpha$'- NaV$_{2}$O$_{5}$ has been identified as another inorganic Spin-Peierls compound
\cite{1}$^{,}$\cite{2} besides CuGeO$_{3}$. X-ray and neutron scattering on polycrystalline material \cite{2},
NMR \cite{3} and single crystalline
susceptibility measurements \cite{4} have been performed on this material. It exhibits a phase transition at
$T_{\rm SP}=35~K$. Below {\it T\/}$_{\rm SP}$ a dimerization is observed in a superlattice reflection which shows
a strong temperature dependence. In addition a gap energy from inelastic neutron scattering of  about 10~meV was deduced
from polycrystalline data. Since magnetic chains (V$^{4+}$) along the b-axis of this orthorhombic compound are surrounded
by nonmagnetic ones  (V$^{ 5+}$) we expect a smaller interchain exchange than in CuGeO$_{3}$.

In this communication we report on magnetic resonance experiments in the frequency range 75-350~GHz for a single crystal
specimen of $\alpha$'-NaV$_{2}$O$_{5}$. We show that we can interpret our data with resonances in the excited triplet state
quantitatively for $T<T_{\rm SP}$ with a singlet-triplet model which reduces to a doublet model for
$T>T_{\rm SP}$.

We used a crystal from the same batch as in ref.~4. It consists of a thin platelet with the chain axis b in the
plane of the platelet. The magnetic field was perpendicular to this plane, parallel to the c-axis.
We used Gunn-oscillators and IMPATT-diodes to generate the mm-wave radiation,
together with frequency multipliers \cite{5}. For our wide frequency range we perform transmission
experiments using oversized waveguides. Since the resonance has a weak absorption strength we used a field
modulation technique. We measured the resonance in Faraday geometry using a superconducting magnet with a field
range up to 14~T.

In fig.~1 we show the field dependence of the measured resonance frequencies from 75 to 330~GHz together with a
display of the signals in the inset for {\it T \/}= 70~K. The linear field dependence gives a {\it g\/}-factor of
$g_{\rm c}= 1.936 \pm 0.002$ below 2 as expected for a V$^{4+}$(3d$^{1}$)-ion.
The {\it g\/}-factor is constant within 0.1~\% in the temperature region investigated of 20~--~100~K.

From the twice integrated signals of fig.~1 we obtain the absorption strength as a function of temperature.
This is exhibited in fig.~2 for a typical selected frequency of 134~GHz.
A very strong increase in the dimerized phase is noted, starting from about 20~K and it reaches a maximum
at about $T_{\rm SP}$ whereas a much weaker temperature dependence is observed in the disordered high
temperature region. These results permit a detailed analysis which could not be done before in CuGeO$_{3}$
because of the smaller $T_{\rm SP}$ and the much faster decrease of the spin gap just below
$T_{\rm SP}$ in the latter substance.

Below $T_{\rm SP}$ we assume  $\Delta q = 0$ transitions within an excited triplet state separated from
the ground state by a temperature dependent gap $\Delta (T)$. We take the temperature dependence of the gap
energy $\Delta (T)$ for $T < T_{\rm SP}$ into account. We take either a molecular field expression
$\Delta (T) = \Delta_{0} (1 - T / T_{\rm SP})^{1/2}$ for the spin gap or we use the temperature
dependence from the experimentally determined superlattice reflection intensity {\it I\/}~~\cite{2}.
If the Spin-Peierls physics is strictly one-dimensional we expect $\Delta \sim \delta$ or $\sim \delta^{2/3}$
with $\delta$ the lattice
distortion. The latter dependence is an exact result \cite{6}. Since the intensity of the superlattice distortion
would go like $I \sim \delta^{2}$ we expect $\Delta \sim I^{1/2}$ or $\sim I^{1/3}$. For the integrated
absorption we get
\begin{equation}\label{one}
  A(T)\sim [\exp \frac{- \Delta + g \mu B}{k_{\rm B}T} - \exp \frac{-\Delta - g\mu B}{k_{\rm B}T}] \mbox{ }/\mbox{ }Z
\end{equation}
where {\it Z\/} is the partition-function involving the singlet and triplet states.
For $T > T_{\rm SP}$ the gap $\Delta =0$ and eq.~(1) leads to excitations within the spin 1/2 doublet.

The full lines in fig.~2 are a fit to the experiment with the parameters
$\Delta(T = 0) = 85$~K (20~K and the corresponding magnetic field (4.9~T) for the frequency of 134~GHz.
The curves are scaled to the experiment at 35~K because of the unknown matrix element and other geometric factors.
A very good agreement with the measurements was obtained for {\it T\/}~$<$~{\it T\/}$_{\rm SP}$ only for
$\Delta \sim$ {\it I\/} (full line). The other two curves $\sim$~{\it I\/}$^{1/3}$ and
~ (1$-${\it T/T\/}$_{\rm SP}$)$^{\frac{1}{2}}$ as well as $\sim$~{\it I\/}$^{1/2}$ (not shown) give poor fits.
This could mean that this substance is not so typically one-dimensional as one usually assumes.
Also fits of eq.~(1) with a temperature independent gap give a poor agreement with the experiment.
Similar results were obtained for 220 and 288~GHz where again a fit with $\Delta \sim I$ gave the best agreement.

For $T > T_{\rm SP}$ the energy spectrum for large fields is rather complicated \cite{7}.
For magnetic fields $B << B_{\rm ex}$ our description should be correct however.
The agreement of the curve based on eq.~(1) with the experiment is good also for $T > T_{\rm SP}$.

The value of $\Delta (T = 0)$ cited above is the one quoted in literature of 85~K for the SP gap \cite{4}
and therefore slightly lower than the gap quoted from inelastic neutron scattering on polycrystalline material
\cite{2}. Since the considered excited triplet can be at any point in {\it k\/}-space
with large enough density of states (preferably at symmetry points) and since there is
appreciable dispersion of the magnetic excitations because of the large
{\it J\/} and $T_{\rm SP}$ our value of the gap of 85~K is quite reasonable \cite{8}.
We expect the gap for the one-dimensional case to occur at $k_{\rm SP} = (0,p,0)$.
One has to await measurements of the full dispersion spectra of magnetic excitations
in order to place our triplet state in $k$-space.

In fig.~3 we show the peak to peak line widths for two resonances (134 and 220~GHz).
The line width is very small compared to similar compounds like CuGeO$_{3}$ \cite{9} or spin 1 systems.
It is typically 40~G wide. This means that in this spin ½ chain there are very few relaxation channels.
In addition the line width has a very weak temperature dependence.
If the line width arises from the spin fluctuations \cite{10} the high magnetic field
suppresses these fluctuations and the absence of any appreciable anisotropy and exchange
anisotropy for the spin ½ chain reduces the line width further.

We also tried to measure ground state excitations by using a far-infrared laser.
In CuGeO$_{3}$ such excitations have been found \cite{11} and they were interpreted as ground
state excitations using a staggered field concept \cite{8} similar to the $S = 1$
compounds NENP, NINO \cite{12}. We did not find any ground state excitations in NaV$_{2}$O$_{5}$ so far.
We hope to detect them with larger single crystals.

Unlike for CuGeO$_{3}$ we could so far only determine the uniform phase U and the dimerized phase
D but not a possible incommensurate phase I. This I-phase would possibly occur for magnetic fields in excess of 16~T,
the limit of our field at present. The absorption strength has a very strong temperature
dependence for $T < T_{\rm SP}$ mirroring the strong temperature dependence of the gap
energy and the large value of the gap. The intensity goes practically to
0 for $T < 20$~K with the exception of 288~GHz.
This is in contrast to the magnetic susceptibility where a large
Van Vleck type term is found at low temperatures.
Comparing the line widths of the two Spin-Peierls compounds, CuGeO$_{3}$ and $\alpha$'-NaV$_{2}$O$_{5}$,
we find for the latter a much weaker temperature dependence indicating probably the more isolated spin chains.

After completion of this work we got notice of similar investigations
on $\alpha$'-NaV$_{2}$O$_{5}$ done at 36.2~GHz \cite{13} and 9.2~GHz \cite{14}.
These authors interpret their data by the susceptibility  formula of  Bulaevskii \cite{15}
with a temperature independent gap, which is strictly correct only for DC
susceptibility measurements at $T << T_{\rm SP}$.
The data of these authors, especially for {\it T} close to $T_{\rm SP}$,
we can fit nicely also with equation (1) using again the temperature dependent gap used for our results.

\acknowledgments

This research is supported by the DFG through SFB 252 and by BMBF.
We thank U. L\"{o}w, M. Sieling and G. Uhrig for fruitful discussions.

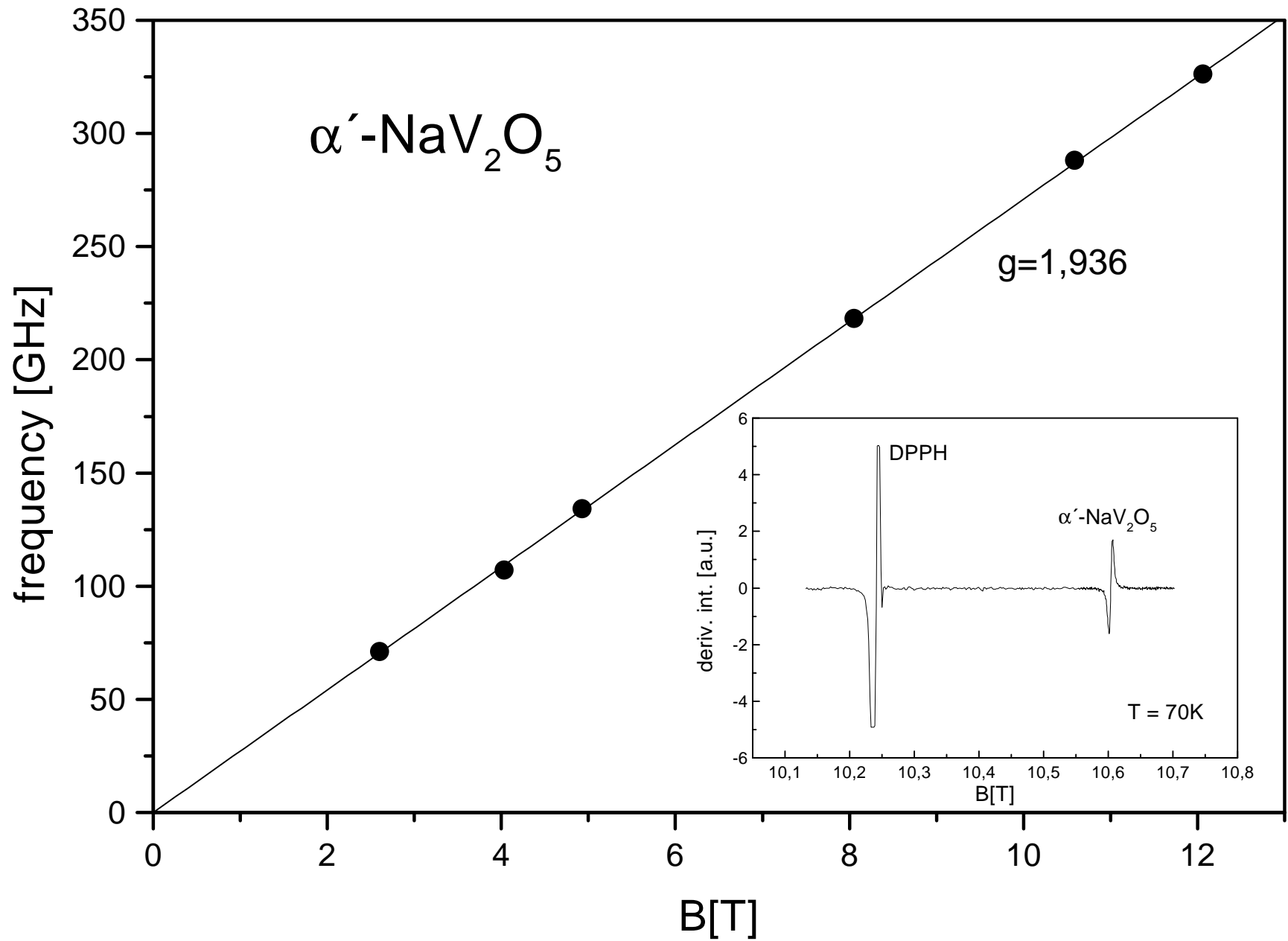
\begin{figure}
\caption{Field dependence of the resonance frequencies from 75 to 330~GHz.
The {\it g\/}-factor is 1.936 for $B$~$||$~c-axis. In the inset the derivative
signals are shown for {\it T =\/} 70 K for DPPH (for field calibration) and $\alpha$'-NaV$_{2}$O$_{5 }$for 288 GHz.}
\label{fig1}
\end{figure}

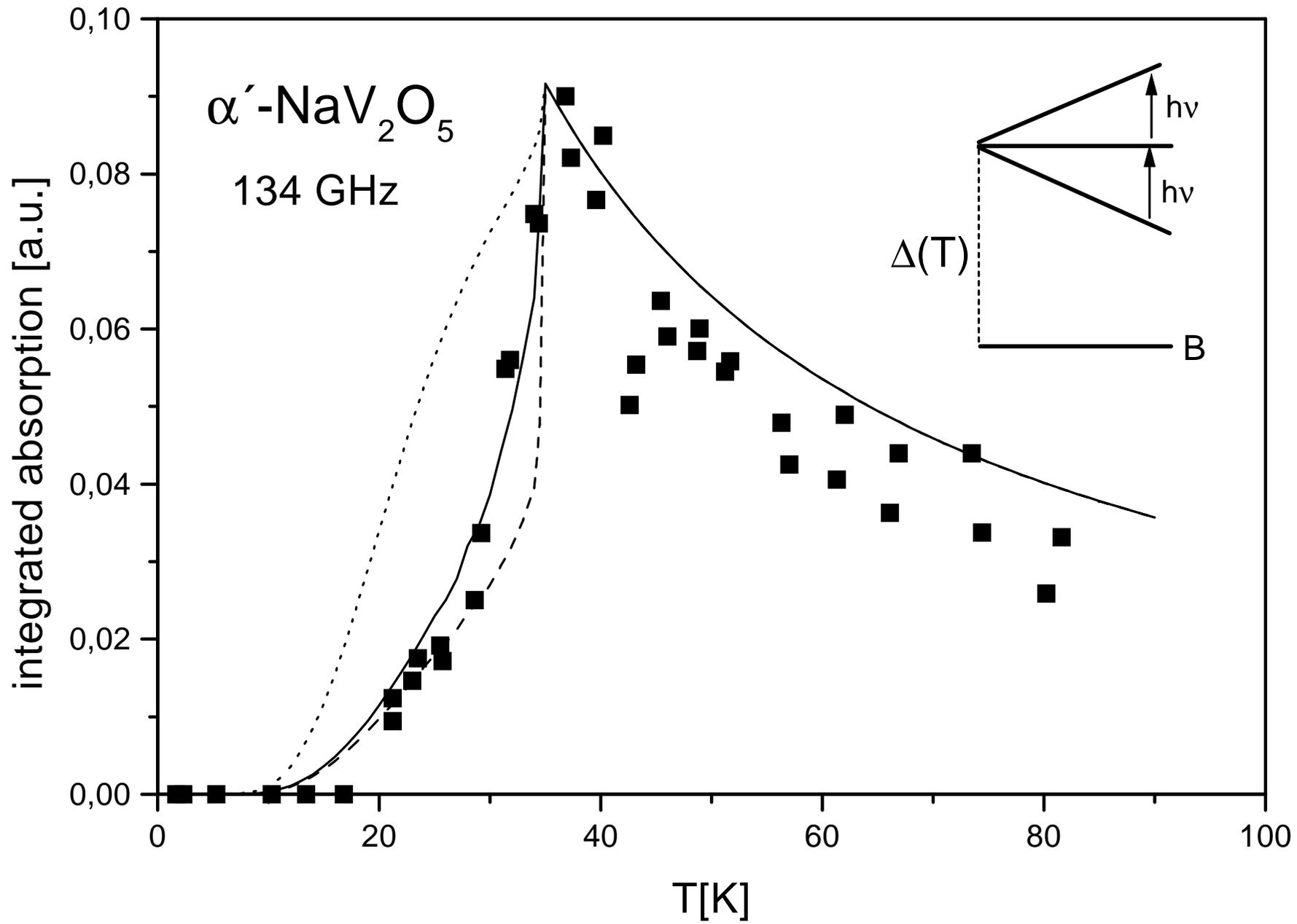
\begin{figure}
\caption{Absorption strength versus temperature for a resonance line at 134~GHz.
The full and the dashed lines are a fit to equation (1) with the gap energy
scaled as described in the text, the dotted curve is a fit using molecular field theory.
In the inset we show the singlet-triplet model.}
\label{fig2}
\end{figure}

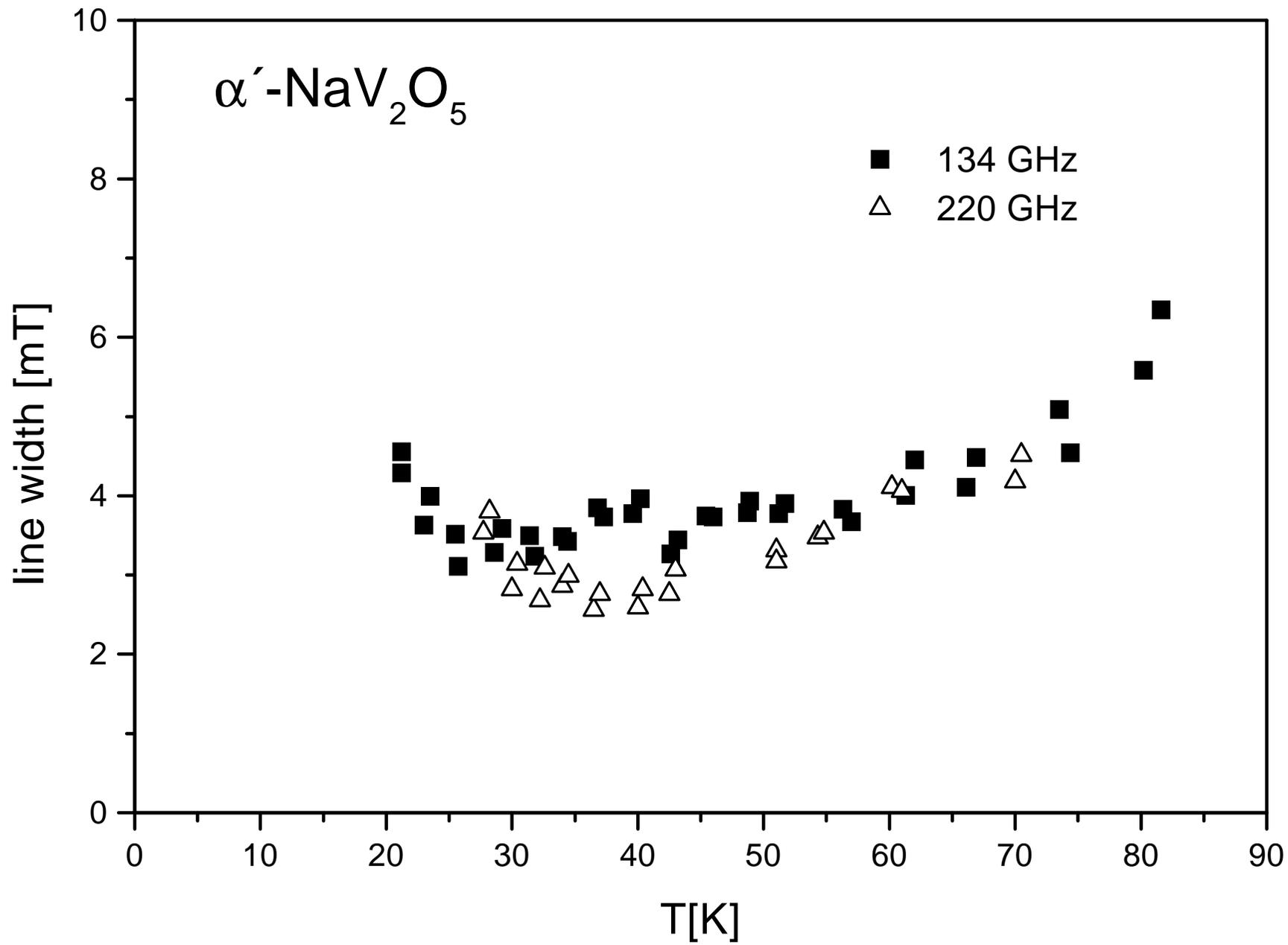
\begin{figure}
\caption{Line width of the resonances as a function of temperature for various frequencies: 134 and 220~GHz.}
\label{fig3}
\end{figure}

\end{document}